\documentclass[sigconf]{acmart}

\AtBeginDocument{%
  }

\copyrightyear{2025}
\acmYear{2025}
\setcopyright{cc}
\setcctype{by}
\acmConference[WWW Companion '25]{Companion Proceedings of the ACM Web Conference 2025}{April 28-May 2, 2025}{Sydney, NSW, Australia}
\acmBooktitle{Companion Proceedings of the ACM Web Conference 2025 (WWW Companion '25), April 28-May 2, 2025, Sydney, NSW, Australia}
\acmDOI{10.1145/3701716.3717526}
\acmISBN{979-8-4007-1331-6/2025/04}

\usepackage{subcaption}
\usepackage[T1]{fontenc}
\usepackage[utf8]{inputenc}
\usepackage{microtype}
\usepackage{makecell}
\usepackage{tabularx}
\usepackage{balance}

\begin{document}

\title{Agent-Initiated Interaction in Phone UI Automation}

\author{Noam Kahlon}
\orcid{0009-0004-0097-7417}
\affiliation{
\institution{Google Research}
\city{Mountain View}
\state{CA}
\country{USA}}
\email{kahlonn@google.com}
\authornote{Corresponding author}

\author{Guy Rom}
\orcid{0009-0004-2537-9018}
\affiliation{
\institution{Google Research}
\city{Mountain View}
\state{CA}
\country{USA}}

\author{Anatoly Efros}
\orcid{0009-0000-1451-3438}
\affiliation{
\institution{Google Research}
\city{Mountain View}
\state{CA}
\country{USA}}
\email{talef@google.com}

\author{Filippo Galgani}
\orcid{0000-0002-3553-3611}
\affiliation{
\institution{Google Research}
\city{Mountain View}
\state{CA}
\country{USA}}
\email{galganif@google.com}

\author{Omri Berkovitch}
\orcid{0009-0009-4337-4954}
\affiliation{
\institution{Google Research}
\city{Mountain View}
\state{CA}
\country{USA}}
\email{berkovitchomri@google.com}

\author{Sapir Caduri}
\orcid{0009-0009-2298-0605}
\affiliation{
\institution{Google Research}
\city{Mountain View}
\state{CA}
\country{USA}}
\email{sapir@google.com}

\author{William E. Bishop}
\orcid{0009-0002-4377-3989}
\affiliation{
\institution{Google Research}
\city{Mountain View}
\state{CA}
\country{USA}}
\email{willbishop@google.com}

\author{Oriana Riva}
\orcid{0009-0007-4514-009X}
\affiliation{
\institution{Google Research}
\city{Mountain View}
\state{CA}
\country{USA}}
\email{oriva@google.com}

\author{Ido Dagan}
\orcid{0009-0007-0165-606X}
\affiliation{
\institution{Google Research}
\city{Mountain View}
\state{CA}
\country{USA}}
\affiliation{%
\institution{Bar-Ilan University}
\city{Ramat Gan}
\country{Israel}}
\email{idodagan@google.com}

\renewcommand{\shortauthors}{Noam Kahlon et al.}

\authorsaddresses{%
Authors' addresses: N. Kahlon, Google, 98 Yigal Alon St, Tel-Aviv, Israel; }

\begin{abstract}

Phone automation agents aim to autonomously perform a given natural-language user request, such as scheduling appointments or booking a hotel. While much research effort has been devoted to screen understanding and action planning, complex tasks often necessitate user interaction for successful completion. Aligning the agent with the user's expectations is crucial for building trust and enabling personalized experiences. This requires the agent to proactively engage the user when necessary, avoiding actions that violate their preferences while refraining from unnecessary questions where a default action is expected. We argue that such subtle agent-initiated interaction with the user deserves focused research attention.

To promote such research, this paper introduces a task formulation for detecting the need for user interaction and generating appropriate messages. We thoroughly define the task, including aspects like interaction timing and the scope of the agent's autonomy. Using this definition, we derived annotation guidelines and created AndroidInteraction, a diverse dataset for the task, leveraging an existing UI automation dataset. We tested several text-based and multimodal baseline models for the task, finding that it is very challenging for current LLMs. We suggest that our task formulation, dataset, baseline models and analysis will be valuable for future UI automation research, specifically in addressing this crucial yet often overlooked aspect of agent-initiated interaction. This work provides a needed foundation to allow personalized agents to properly engage the user when needed, within the context of phone UI automation.

\end{abstract}



\maketitle

\section{Introduction}

Autonomous agents capable of interacting with a smartphone's graphical user interface (GUI) to complete tasks on behalf of users have drawn significant interest \cite{yan2023gpt, wen2023empowering, yang2023appagent, li2024personal}. The agent, presented with a natural language (NL) instruction and a starting screen, iteratively executes the instruction through a series of GUI interactions. The agent can perceive the screen, potentially accessing UI metadata such as the accessibility tree,\footnote{An accessibility tree is a hierarchical structured representation of the UI elements in a screen, providing information about their type, properties, and relationships. It is often used by assistive and other technologies to facilitate interaction with the UI.} and is capable of performing a wide range of GUI actions, such as tapping, swiping, and inputting text. These agents hold the promise of substantially increasing user experience and efficiency in phone-based tasks and improving accessibility for individuals who are physically or situationally impaired \cite{li2023uinav, li2024personal, wang2023enabling}.

To truly personalize these agents and enable them to effectively serve individual users, a deeper understanding of user needs and preferences is crucial. This necessitates moving beyond simple task execution and empowering agents to actively engage in interactions with users. By proactively seeking information or clarification, these agents can tailor their actions to individual preferences and avoid unintended consequences.
This introduces a critical need for alignment between the agent's actions and the user's intentions, which is essential for fostering user trust and creating personalized experiences. As a step in this direction, our work explores the yet underexplored area of agent-user alignment in phone automation, particularly in scenarios where interaction with the user is needed.

\begin{figure}[t]
    \centering
    \includegraphics[width=\columnwidth]{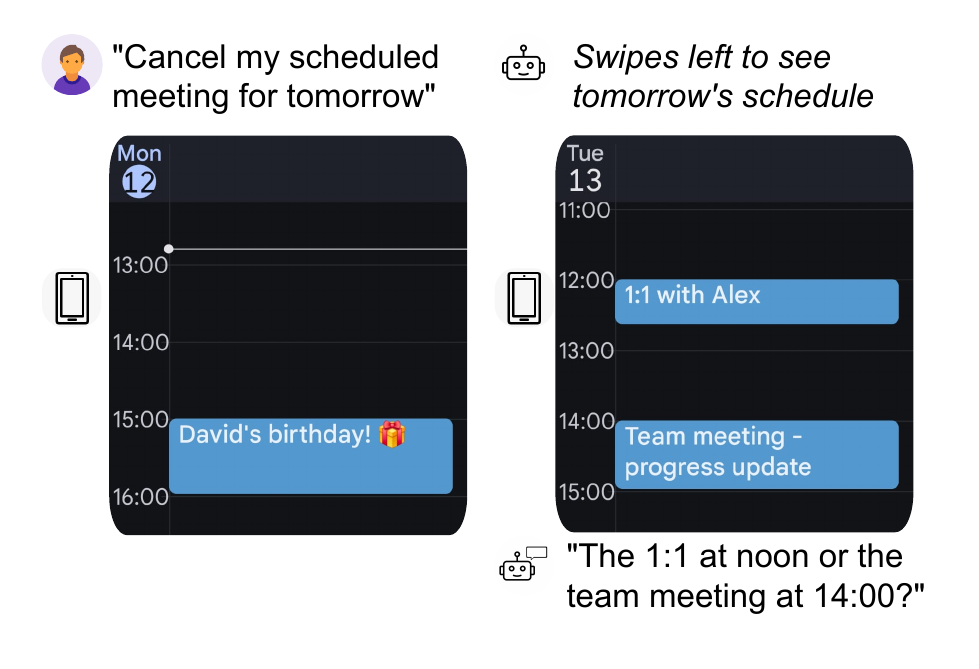}
    \caption{An illustrating example of an instruction requiring user interaction in phone UI automation. The user requests to cancel their meeting tomorrow, unaware that there are two meetings scheduled for the next day. See figure \ref{fig:scratch_episode} and appendix \ref{app:additional_examples} for additional examples.}
    \Description{Screenshots from a phone's Calendar app, illustrating a scenario where a user has requested to cancel their meeting tomorrow. The last screenshot displays the next day's calendar with two meetings scheduled on different times. This necessitates the agent formulating a message to specify which meeting they want to cancel.}
    \label{fig:motivation_episode}
\end{figure}

In fact, many tasks necessitate interaction with the user for purposes such as gathering additional information, confirming sensitive actions or clarifying ambiguous instructions \cite{wang2023enabling, sun2022meta, todi2021conversations}. For instance, given an initial screen displaying a document, the instruction "send this doc to Alex" could trigger several types of user interactions. The agent might ask the user to clarify which contact named Alex to send it to or which app to use. If the document is sensitive, the agent might seek confirmation before proceeding. The agent might also need to inform that the task is infeasible, for example if the document is locked for sharing (See Figure \ref{fig:motivation_episode} for an additional illustrated example). 
In many cases, particularly with complex tasks, the need for user interaction arises dynamically during task execution \cite{lu2024weblinx}, making its detection crucial to maintain user control and ensure alignment between the agent's actions and the user's preferences \cite{kim2024aligning}. Despite its importance, this aspect remains largely unaddressed in current UI automation research (see Section 2.2 for further discussion).

An autonomous agent capable of user interaction must be able to: 1) detect the need for interaction, 2) formulate an appropriate message to the user, and 3) integrate the user's response to proceed with the task. This paper focuses on the first two challenges. As an initial step towards tackling these challenges, we leverage an existing phone automation dataset, which did not originally include user interaction, to develop AndroidInteraction, an effective dataset for detecting the need for user interaction and generating appropriate messages\footnote{The dataset is available at \url{https://github.com/google-research/google-research/tree/master/android_interaction}}. This involved a role-playing exercise where annotators simulated user interactions to capture diverse needs and preferences. AndroidInteraction embodies several inherent challenges of this task, such as the subjectivity in assessing when interaction is required. We address these challenges in our task formulation and data collection process. Additionally, we present and evaluate several baseline models on the AndroidInteraction dataset, and analyze their performance. We believe our work can facilitate research on agent-initiated interaction within the realm of phone UI automation.

Overall, we make the following contributions:    
\begin{itemize}
    \item Introducing a novel task formulation for agent-initiated interaction in phone UI automation, including detecting the need for interaction and generating a suitable message to the user.
    \item Creating AndroidInteraction, a diverse dataset, of over 750 demonstrations spanning over 250 different apps, which allows to evaluate models' performance on the interaction task. 
    \item Introducing, evaluating and analyzing several baseline models for the task using our dataset.
\end{itemize}

\section{Background} 

\subsection{UI Automation Agents} \label{ui_automadialtion_bg} 

Phone automation agents aim to autonomously complete tasks on behalf of the user. 
A prevalent approach involves interacting with apps via APIs, similar to slot-filling in task-oriented dialogue systems \cite{zhan2023you, wen2023empowering}, but this method is constrained by the need for application-specific API access and manual configuration of task workflows.
In contrast, UI-based task automation simulates human-UI interactions, offering greater generality and adaptability but facing challenges in generalizing to complex tasks across diverse apps \cite{yang2023appagent, wang2023enabling, sun2022meta}.

In order to represent the UI for UI-based automation, a primary approach uses textual formats \cite{wang2021screen2words, zhang2021screen}, such as OCR and icon labels \cite{rawles2023android}, or system-provided accessibility trees \cite{bishop2024latent}. Another approach utilizes multimodal models \cite{yan2023gpt, hong2023cogagent} to integrate visual information from the screen. While the latter approach incorporates valuable visual cues, it faces challenges in data requirements and complexity \cite{yang2023appagent}.


UI automation agent evaluation typically falls into two categories. One approach \cite{toyama2021androidenv, schneider2022mobile} involves simulating an interactive phone environment, typically measuring task fulfillment success. This approach allows a detailed analysis of efficiency and quality, but are harder to develop across diverse apps and tasks.

A more common evaluation method utilizes static gold datasets \cite{li2020mapping, venkatesh2022ugif, burns2022dataset}, composed of recorded human demonstrations of task execution. Each recording is a sequence of action-observation pairs, where observations include screenshots and metadata (like icon recognition or OCR), along with the annotator's actions. Agent evaluation involves step-wise comparison of its selected actions with the recorded actions. However, measuring accuracy in this approach is challenging as there are almost always multiple ways to execute the same user request. For example, the instruction "delete my last email sent to David" could be accomplished by either searching for "David" or navigating to the sent emails folder.

\begin{figure*}[t]
    \centering
    \includegraphics[width=\textwidth,height=6cm]{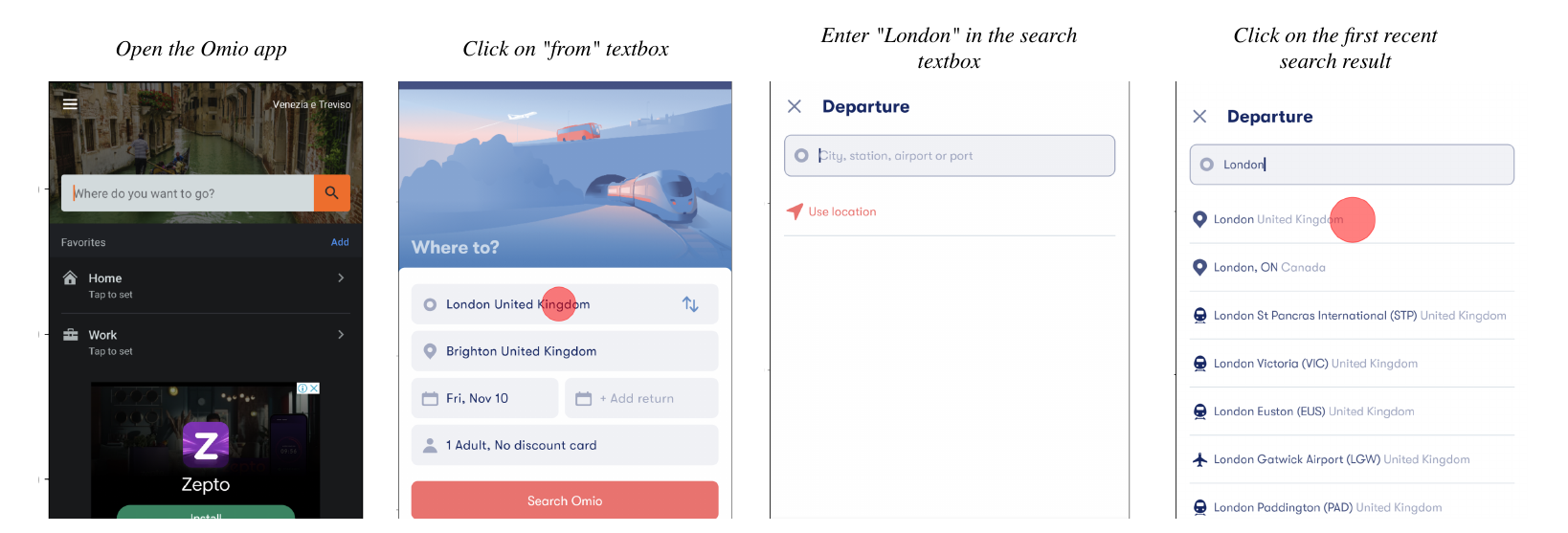}
    \caption{First steps of an example episode from the AndroidControl dataset. The user instruction is to book a train from London to Brighton on a specific date using the Omio app.
    The 4th step was labelled as requiring user interaction in our dataset, with the message: "From which London station would you like to depart?".}
    \Description{Four phone screenshots illustrating an automated task of booking a train ride from London to Brighton. The fourth screenshot displays multiple possible departure stations that match the "London" search query, suggesting a needed question from the agent to clarify the user's intention.}
    \label{fig:scratch_episode}
\end{figure*}

\subsection{User Interaction in UI Automation}  \label{ui_interaction_bg}

The need for user interaction for task completion has long been acknowledged within traditional task-oriented dialogue systems. These systems typically operate within a restricted domain and rely on a predefined data schema, with dialogues primarily serving to pull information for slot-filling \cite{zhang2020recent, chen2017survey}. 
Conversely, phone UI automation operates in a \textit{generic}, open-domain setting, presenting additional challenges due to the need to proactively interact with a dynamic and visually rich environment. User interaction may be necessary not only for disambiguating intents or filling in missing details but also for addressing unforeseen situations arising from UI interactions, such as confirming potentially sensitive actions. 
Despite this, most datasets \cite{rawles2023android} and models \cite{zhan2023you} for phone UI automation neglect the user interaction aspect, and assume that tasks can be executed without user involvement.

Some exceptions do exist, though. \citet{wang2023enabling} proposed a categorization of user interaction based on two binary dimensions, resulting in four distinct scenarios: the initiator --- user or agent, and the purpose --- soliciting or providing information. However, their data concerning agent-initiated interaction is limited. The portion addressing information solicitation only deals with clarification questions about empty fields on a single screen. The portion dealing with information providing is restricted to single-screen summarization. Both represent only a fraction of the agent-initiated interactions necessary for complex task completion.
\citet{sun2022meta} introduced a model and dataset for phone UI automation tasks involving multi-turn conversations. Yet, their data focuses on a limited set of apps and interactions, lacking diversity.  \citet{burns2022dataset} focused on a specific setting of detecting task infeasibility, by examining feasible instructions on one app that were infeasible on a similar app. Lastly, \citet{lu2024weblinx} presented recently a conversational-agent dataset and model, in the domain of web agents, which is closest to our targeted scenario. 
This dataset is complementary to ours in that it covers web browser interactions rather than phone apps automation, and in addressing interaction mainly within smaller sub-tasks that are specified by the user during a longer interactive browser session.

\subsection{The AndroidControl Dataset} \label{scratch_desc}

For our research, we leverage a subset of the data collecte for AndroidControl \cite{li2025effects}, an existing crowdsourced dataset for UI automation. Annotators were given generic task descriptions and asked to instantiate them into specific instructions, then record their execution on actual phone using an app of their choice. The dataset covers diverse app categories, including shopping, travel, education, calendaring, file management, and more.

Each task recording, referred to as \textit{episode}, contains a high-level NL instruction provided by the annotator and a sequence of \textit{steps}. Every step consists of an observation (a screenshot and its accessibility tree), and comprehensive information about the recorded action: the (x,y) position(s) for touches or swipes, any typed text, and an NL description of the action (e.g., "scroll down," "click on the search bar"). Figure \ref{fig:scratch_episode} shows a portion of an example episode.

A key feature of AndroidControl is the diversity of task instructions and execution paths, aiming to emulate real-world scenarios. 
Additionally, many of the instructions are based on LLM-generated persona profiles, with elaborate fake personal details such as email and home addresses, family members and hobbies and information like planned vacations, day-to-day schedule, etc.

\section{User Interaction in UI Automation: Task Definition} \label{task_def}
Our task is defined within a phone UI automation setup (see Section \ref{scratch_desc}). In our setting, each session begins with a natural language user instruction and a starting screen. The agent attempts to execute the instruction step-by-step through actions on the UI. However, unlike other common UI automation settings, we assume that the agent can contact the user and refer to contents of the screen, which the user can see, in order to complete the current session.

Our task asks to determine, at any step of the session, whether the agent should contact the user to successfully complete the session, given the instruction and session history so far. If an interaction is required, the task further requires formulating a clear message that effectively conveys the necessary information or request to the user.

We introduce the following assumptions to establish our intended scope for this task, which was further captured in our annotation guidelines:

\paragraph{Scope of the User Instruction:} We limit the session's scope to the given user instruction and to the interactions arising directly from its execution. Specifically:
\begin{itemize}
    \item The agent will not suggest alternative tasks to the original instruction. For example, if the user asks "where is the nearest Walmart?" and the maps application shows a Target store nearby, the agent will not suggest going there instead. However, if the app itself raises this question - the agent should deliver the question to the user.
    \item The agent has no internal policy-based restrictions on the types of instructions it can execute, and is accordingly willing to perform any instruction given by the user. For example, such a policy might force the agent to contact the user before carrying out any payments,
    but those policy-based interactions are excluded in our setting.
    \item The agent is assumed to be proficient in UI navigation and capable of executing any feasible instruction. Specifically, the agent will not ask the user for help navigating the UI.
\end{itemize}

\paragraph{Interaction Timing:} The agent will contact the user when interaction is needed for its next action. If there are multiple possible actions on the screen (e.g., several empty fields) we assume the agent processes these sequentially from top to bottom, mimicking typical user behavior. This ensures the interactions are timely, and their timing is well defined. We avoid prematurely anticipating future needs or unnecessarily delaying essential interactions. For example, given the user instruction "Order me a pair of size 44 shoes" the agent will not ask "Of which brand?" until it reaches the brand selection option in its workflow. On the other hand, once the agent encounters that option, it will ask promptly for the brand rather than postponing the interaction to fill in the shoe size, for example.



\paragraph{User preferences:} A key aspect of user preferences lies in how much autonomy a user wishes to delegate to the agent and how tolerant they are to the agent's requests for information. For instance, a user instructing the agent to buy a specific shirt might be receptive to a website's offer for a discount on two shirts, while another user might find it intrusive. Another common scenario involves fields that were not specified in the instruction and can be verified with the user or left with their default value. By capturing a spectrum of necessity levels for interactions, our data provides insights into varying user preferences and tolerances, which can be leveraged to develop agents that adapt to individual needs. We further address these considerations in our data collection methodology (see Section \ref{data_subjectivity}).


\section{AndroidInteraction}

To create AndroidInteraction, we augmented a subset of the episodes collected for the AndroidControl dataset (see Section \ref{scratch_desc}) with annotations indicating where interactions are required, along with the appropriate messages to present to the user. This section presents the structure of our dataset records, the annotation methodology employed, and the considerations we made to ensure data quality while addressing the inherent subjectivity of user preferences.

\subsection{Dataset Structure} \label{record_structure}
For each step in an AndroidControl episode, we annotated whether a user interaction is needed at that step. Where relevant, the annotator provided a corresponding message that asks for the needed user response, as well as a 1-5 score for the expected necessity of user interaction. The necessity score reflects whether the interaction is strictly essential for task completion or involves a preference that could potentially be handled by the agent, e.g. with a default value. This score was determined while considering potential different preferences of different users, aiming to capture the subjectivity of such preferences. See Table \ref{table:necessity_examples} for examples of interactions with different necessity levels, and further discussion about assessing necessity level in the next section.


\begin{table*}
\centering
\caption{Examples of interactions annotated with different necessity levels, ranging from 5 (essential for the next action) to 1 (may annoy some of the users)}
\begin{tabularx}{\linewidth}{X p{4.5cm} p{7cm} p{4cm}} 
\toprule
\textbf{Level} & \textbf{User instruction} & \textbf{Episode description} & \textbf{Message} \\ 
\midrule
 5 & Identify a plant using a picture taken today & Upon opening the phone gallery there are several suitable images of plants & "Which picture would you like to use?" \\
 5 & Change my password in the MuniMobile app & Opening the app and entering security settings & "Please enter your current password" \\
 3 & Turn on the "Bump tracker" in the Ovia Pregnancy app & When the app is opened, a pop-up appears requesting permission to send notifications. & "Allow the Ovia app to send you notifications?"\\
 3 & Share this document with a specific person via GMail & The "share" action generates an email with the document attached, but leaves the subject line blank. & "Would you like to add a subject to the email?"\\
 1 & Download a specific file from Google Drive & Navigating to the file and clicking download opens the "Downloads" folder as the default destination & "Would you like to use this folder?" \\ 
\bottomrule
\end{tabularx}
\label{table:necessity_examples}
\end{table*}

\subsection{Data Collection Methodology} \label{data_collection_method}

AndroidInteraction was annotated by 2 outsourced professional annotators, after thorough guidance and several pilot sessions. Each annotator independently labeled a little over 600 unique episodes. We further sampled 200 of the first annotator's episodes to be relabelled by the second annotator, to evaluate inter-annotator agreement.

Detailed instructions guided the annotators to envision themselves as human agents acting for users who cannot directly operate their phone, similar to a remote desktop support. This role-playing approach allowed us to effectively simulate the role of an automated UI agent and capture diverse preferences in a realistic setting. They were to assume no prior familiarity with the user, and rely solely on the user's instruction, screenshots and session history. The annotators were asked to identify episode steps where user interaction was necessary to complete the current session, and to compose a clear message allowing the user to easily resolve the situation. The full annotation guidelines will be released with the publication of this paper. Note that following our task scope, as defined in the previous section, interactions confirming task completion, requesting new tasks, or suggesting further actions were not collected.


\paragraph{Data Selection:}   

We applied several filtering criteria to ensure data quality. First, we excluded sessions where the original AndroidControl annotator indicated an unsuccessful outcome due to reasons unrelated to the need for user interaction (e.g., technical issues, annotator confusion). Second, when our annotators identified a necessary interaction, we truncated the session at that point. This maintained consistency, as subsequent steps in the original episode rely on arbitrary choices made by the annotator, in the absence of user input. Consequently, all the sessions in AndroidInteraction contain at most one interaction, a limitation further discussed in Section \ref{dataset_limitations}. Additionally, we removed records potentially containing private information to respect user privacy. Finally, we aimed to filter out `open' tasks, where instructions were intentionally vague. Such tasks often necessitate an ongoing dialogue with the user to achieve satisfactory completion, a scenario beyond our current scope. For example, asking the agent to browse a category of products while aiming to find relevant candidates would require a conversational dialogue, which is of a different nature than UI automation that is typically mostly autonomous. After applying these filtering steps, our final dataset comprises 772 episodes.



\paragraph{Assessing Necessity Level:} \label{data_subjectivity}
While certain interactions are strictly necessary for task completion, the agent can often infer a reasonable default choice and proceed without interrupting the user. However, relying solely on default values might not align with the user's intent. For instance, given the instruction "set a timer", it is sensible to ask "for how long?" even if the app suggests a default duration. Yet, asking whether the user wishes to name the timer is unreasonable, even when such option exists in UI. This, together with varying user tolerance for interruptions, introduces subjectivity in determining when an interaction is truly justified. Instead of strictly classifying interactions as 'essential' or 'non-essential', we acknowledged a spectrum of necessity, with each interaction in our dataset assigned a 1-5 score reflecting its level of necessity based on our guidelines. This nuanced approach to necessity not only reflects the diversity of user preferences but also provides valuable data to training agents with personalized interactions that generate contextually appropriate messages.

\subsection{Data Quality Analysis} \label{data_quality_assessment}

As mentioned above, 200 episodes were annotated by both annotators, for inter-annotator agreement analysis. 
With respect to determining whether an interaction is required in each step, inter-annotator agreement achieved a Cohen's Kappa score of 0.64, considered as moderate agreement, which can be expected given the inherent subjectivity of the task. 

With respect to the messages generated by the annotators, we manually analyzed all the steps labelled as requiring interaction by both annotators and found that their messages were semantically equivalent (paraphrastic), in the context of the UI session, in all the cases.
Further, there was a high correlation (0.78 Pearson) between their assigned necessity scores.

In addition, to analyze and assure message quality, we manually reviewed all messages generated by the annotators. We found them almost always appropriate with less than 5\% requiring minor stylistic edits for grammar, punctuation, or clarity, which we incorporated into the data in a final pass.


\subsection{Dataset Analysis}

\begin{figure}[htb]
    \begin{subfigure}{0.475\columnwidth}
        \includegraphics[width=\columnwidth]{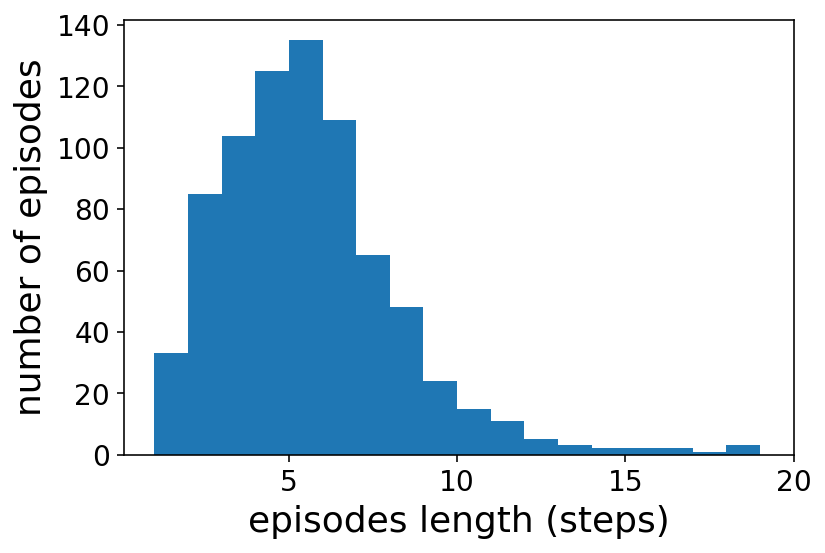}
        \caption{}
        \label{fig:length_dist}
    \end{subfigure}
    \begin{subfigure}{0.475\columnwidth}
        \includegraphics[width=\columnwidth]{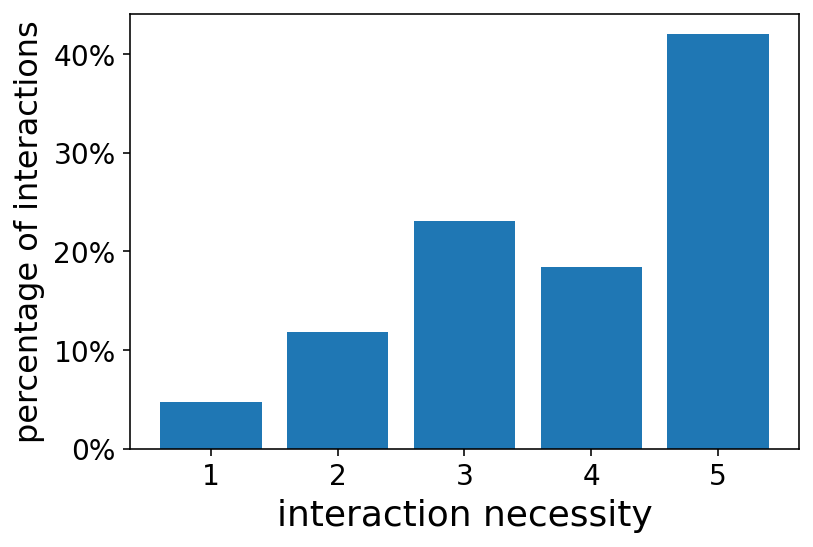}
        \caption{}
        \label{fig:necessity_dist}
    \end{subfigure}
    \begin{subfigure}{\columnwidth}
        \centering
        \includegraphics[width=0.6\columnwidth]{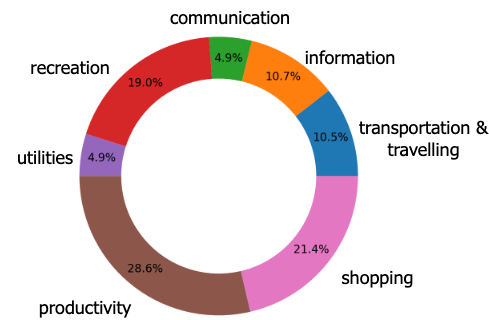}
        \caption{}
        \label{fig:app_domains}
    \end{subfigure}\\[-1ex]
    \caption{Distribution of (a) the episodes length and (b) the interactions necessity scores in AndroidInteraction. (c) shows the distribution of the domains for the various apps in the dataset}
    \Description{(a) The distribution of episode lengths in AndroidInteraction, which exhibits a unimodal right-skewed distribution, with most episodes between 3 and 7 steps long; (b) a bar chart of the percentage of records for each level of interaction necessity, showing an increasing trend with about 40\% at level 5; (c) a donut chart of the various domains of apps in the dataset. "productivity" and "shopping" account for almost 50\% combined, and the remaining apps are spread across "recreation", "information", "transportation \& travelling", "communications", and "utilities".}
    \label{fig:dataset_stats}
\end{figure}

Our final dataset comprises 772 episodes, consisting of 3,605 steps, with an average episode length of 4.7 steps and length distribution shown in Figure \ref{fig:length_dist}. The instructions span more than 250 different apps, from various domains (see distribution in Figure \ref{fig:app_domains}).
Interactions were observed in 27.5\% of the episodes, resulting in 6\% of the steps annotated as requiring an interaction. The dataset was split into a 70\% test set and a 30\% validation set. Roughly 65\% of the episodes were collected using simulated personas (see Section \ref{scratch_desc}), from a pool of 17 personas, promoting data diversity. The distribution of necessity scores (Figure \ref{fig:necessity_dist}) shows the annotators labelled more interactions with higher necessity, particularly those deemed essential (necessity score 5).


\paragraph{Types of User Interaction}

Our task focuses on interactions required for the agent's next action, which  represent 3 main scenarios: (1) multiple options exist for the next action (e.g., due to unspecified choices or missing fields), triggering the agent to seek needed information from the user; (2) no feasible action exists to complete the task, leading the agent to report failure; (3) a clear next action exists, but the agent seeks confirmation from the user before proceeding, to avoid potential harm. Recall that in our task definition (Section \ref{task_def}) we assume agent proficiency in UI navigation, hence exclude navigation-related questions or failures. Thus, failure messages are limited to cases where the task infeasibility is evident from the screen. 

To gain insights into the distribution of these interaction types, we manually analyzed a sample of 100 interactions. As shown in Table \ref{table:dataset_categories}, the majority (89\%) were requests for more information. Confirmations accounted for 7\%, primarily in cases where the next action involved actions impacting user privacy (e.g., registration, location sharing). Infeasible tasks constituted 4\% of the sample. Among requests for information, approximately 80\% of them involved choosing a value from a closed set of options. Typical examples include selecting a product size, choosing a file format, or specifying a particular train station or airport when only a city name was provided in the instruction. The remaining information requests pertained to less structured choices, such as filling open-text fields (e.g. personal details, email subject). Notably, 20\% of all interactions occurred at step 0, indicating that further information was needed before the agent even began executing the instruction. 






\begin{table}[ht!]
\centering
\caption{Analysis of interaction types.}
\begin{tabular}{l c} 
 \toprule
 \textbf{Category} & \textbf{Percentage} \\ 
 \midrule
 infeasible & 4\% \\ 
 confirmation & 7\% \\
 more information & 89\% \\
 \hspace{3mm}from a closed set & 74\% \\
 \hspace{3mm}open-text field & 15\% \\
  \bottomrule
\end{tabular}
\label{table:dataset_categories}
\end{table}


\subsection{Dataset Challenges and Limitations} \label{dataset_limitations}
Leveraging the existing AndroidControl dataset allowed us to annotate a diverse range of interactions across various apps and scenarios. However, since AndroidControl did not originally include user interactions, our data addresses only the first needed interaction in a session, not covering multi-turn interactions. While this allows us to address some foundational aspects of the problem --- detecting the need for interaction and phrasing adequate messages --- it leaves the multi-turn aspects unaddressed. Evaluating episodes with multi-turn interactions will require additional metrics, such as successful task completion in accordance with user replies. This also opens the possibility of expanding the dataset with more complex, open-ended tasks that require  back-and-forth dialogue to resolve.


Future work could build upon our initial efforts and address these limitations by employing a more comprehensive data collection methodology that allows for concurrently annotating both the demonstration of UI automation (as was done in the original AndroidControl dataset) and agent-initiated interactions (as was done in our work). As discussed in Section \ref{ui_interaction_bg}, there are only a few limited prior datasets of this kind, which do not cover the diverse scope of our interaction setting and types.

\section{Baseline Models}
We conducted a series of experiments to assess the effectiveness of common modeling approaches in detecting the need for user interaction and generating appropriate messages. We use F1 score over individual steps to evaluate interaction detection, while message quality was assessed by similarity to the gold message, both manually and using automatic tools. \footnote{Because true gold interactions comprise only ~5.5\% of AndroidInteraction, a model that always predicts "True" will achieve an F1 score of ~0.1. A model randomly guessing labels according to the class distribution would achieve an F1 score of ~0.055.}

Our initial experiments focused on text-based models utilizing the accessibility tree for screen representation, employing a state-of-the-art Gemini 1.5 model \cite{reid2024gemini}. Given the complexity of the raw accessibility tree and irrelevance of many of its elements to our task \cite{wen2023empowering}, we streamlined it, following \cite{bishop2024latent} --- removing irrelevant elements and containers, and formatting the rest into a hierarchical list with natural language descriptions. Past agent actions were provided as contextual information to the model. 

In addition to trying common prompting techniques (e.g. chain-of-thought \cite{wei2022chain}, few-shot prompting), we primarily focused on a two-stage architecture. The first stage, similar to other approaches in planning and user interaction \cite{deng2023mind2web, ren2023robots}, first listed plausible agent actions.
The second stage utilized these plausible actions and a detailed description of the task in order to determine whether user interaction was necessary and to phrase an appropriate message if so. All experiments were conducted with both zero-shot and few-shot prompts, which will be released with the publication of this paper. Due to the high rate of false positives in most models (see Table \ref{table:model_results}), we attempted to add an audit component to filter out false positives, but this did not yield significant performance improvements.


In addition to the text-based approach, we explored a multimodal approach  
by incorporating screenshots into the Gemini 1.5 model's input. We evaluated the model's performance using screenshots alone, as well as in combination with the accessibility trees. The same prompts developed for the text-only models were repurposed for both single-stage and two-stage architectures in this multimodal setting.

\section{Results and Analysis} \label{experiments}

The results of our experiments, with respect to detecting the need for user interaction, are summarized in Table \ref{table:model_results}, indicating that the models struggle with this task. This is primarily evident in the low precision scores, suggesting a tendency towards false positives. An examination of these false positives reveals that the models often adhere rigidly to instructions, resulting in unnecessary confirmations and navigation questions (see error analysis in Appendix \ref{app:error_analysis}). Notably, the two-stage approach outperforms the single-stage one, and few-shot prompting outperforms zero-shot prompting on F1, improving precision at the cost of recall and yielding more balanced performance. We hypothesize that the overall limited performance of foundation models on this task stems from the limited prevalence of relevant data for this specific setting in their pre-training data. Furthermore, the relatively minor improvement from few-shot prompting suggests that the task's nuances might be too subtle to capture effectively with just a few examples. This may indicate that fine-tuning on a larger dataset specifically tailored to this task might be necessary for substantial performance gains.

\begin{table}[tbh!]
\centering
\caption{Model performance on the user interaction detection task, using different input modalities (text: Accessibility tree, scrn: screenshot), model architectures (1-stage, 2-stage), and prompting strategies (Zero-Shot (ZS), Few-Shot (FS)).}
\begin{tabular}{l c c c} 
 \toprule
 & \textbf{precision} & \textbf{recall} & \textbf{F1} \\ [0.5ex] 
 \midrule
 text 1-stage ZS & 0.08 & \textbf{0.84} & 0.15 \\ 
 text 1-stage FS & 0.12 & 0.65 & 0.2 \\ 
 text 2-stage ZS & 0.1 & 0.58 & 0.17 \\ 
 text 2-stage FS & \textbf{0.19} & 0.35 & \textbf{0.25} \\ 
 \midrule
 scrn 1-stage ZS & 0.11 & 0.63 & 0.19 \\
 scrn 1-stage FS & 0.13 & 0.53 & 0.21 \\
 scrn 2-stage ZS & 0.11 & 0.61 & 0.19 \\
 scrn 2-stage FS & \textbf{0.19} & 0.3 & 0.23 \\
 \midrule
 scrn+text 1-stage ZS & 0.09 & 0.63 & 0.16 \\
 scrn+text 1-stage FS & 0.14 & 0.43 & 0.22 \\
 scrn+text 2-stage ZS & 0.12 & 0.60 & 0.20 \\
 scrn+text 2-stage FS & 0.16 & 0.62 & \textbf{0.25} \\
 \bottomrule
\end{tabular}
\label{table:model_results}
\end{table}

Regarding input modalities, F1 scores did not vary significantly across the three modalities (text-only, image-only, and combined). As expected, the combined input of text and image yields the best results, striking a superior balance between precision and recall from a user perspective. It notably gains substantial recall while only sacrificing a small amount of precision compared to the text-only modality. Furthermore, the combined approach leads to improved message quality, as detailed below. Error analysis (Appendix \ref{app:error_analysis}) reveals that many errors in the text-only modality stem from screen understanding issues, a problem that is largely mitigated by incorporating visual input.

For our top performing model, we assessed message quality by examining all the true positives in the detection task. First, one of the authors manually scored each message as equivalent or not to the ground truth message. Additionally, we automatically performed the same scoring by prompting GPT-4o\footnote{https://platform.openai.com/docs/models/gpt-4o} to compare the ground truth message with the model's output, given the user instruction and the relevant screenshot. It is important to note that message adequacy is not solely based on semantic equivalence, but also on contextual appropriateness within the episode. For instance, an agent instructed to cancel an appointment with two viable options (Figure \ref{fig:motivation_episode}) could appropriately ask either "the team meeting or the 1:1?" or "the one at noon or at 14:00?". Table \ref{table:msg_quality} demonstrates that most generated messages were deemed adequate and suitable, particularly those from models utilizing the image modality. Also, the automatic assessment, while suitable for guiding development, struggled to accurately evaluate message quality. Specifically, when compared with human judgement it exhibited a tendency towards excessive criticism, often failing to recognize messages expressing similar intent through different phrasing.

\begin{table}[tbh!]
\centering
\caption{Accuracy of the generated user message for the true positives of our top-performing models, by their input type, assessed manually and using GPT-4o. All models have a 2-stage architecture and used few-shot prompting, as indicated in Table \ref{table:model_results}.}
\begin{tabular}{l c c} 
 \toprule
 & \textbf{manual} & \textbf{GPT} \\ [0.5ex] 
 \midrule
 text input & 74\% & 70.4\%\\ 
 screenshot input & \textbf{91.3\%}  & \textbf{80.4\%} \\
 text+screenshot input & 88.3\% & 74.7\% \\
 \bottomrule
\end{tabular}
\label{table:msg_quality}
\end{table}

\section{Conclusion and Future Work} \label{future_work}
In this paper, we introduced a novel task formulation for phone UI automation agents, focusing on detecting the need for user interaction and generating appropriate messages to facilitate it. To promote investigation of this aspect of UI automation, we created AndroidInteraction, a diverse dataset for the task and evaluated several text-based and multimodal baseline models. Our findings suggest that the task is nuanced and subtle, and we provided insights into prominent dataset properties and model error types. We conjecture that fine-tuning models on a larger dataset may be necessary to achieve optimal performance. We suggest that our work contributes to the development of more capable and user-friendly phone UI automation agents, highlighting the importance of optimal user interaction when performing complex tasks. 


Future work may develop a data collection approach to simultaneously annotate both agent action demonstrations and user interactions in multiple turns, which may be achieved, for example, by leveraging a Wizard-of-Oz (WoZ) paradigm.
Such a WoZ paradigm may also enable direct assessment of message quality, with the "user" annotator evaluating the relevance and clarity of the "agent" messages, offering valuable insights for developing interaction strategies.

Additionally, incorporating memory components 
could further enhance dataset realism and agent performance. By leveraging past interactions or user profiles, an agent could preemptively fill in personal details or infer preferences, potentially reducing the need for explicit clarification. Furthermore, memory-enabled agents could adapt their interaction style to that of the individual user over time.

\section*{Limitations} \label{limitations}

As described in Section \ref{dataset_limitations}, our work has several limitations. First, our dataset is limited to single-turn interactions, as it was derived from an existing dataset that did not include user interactions. This excludes multi-turn dialogues and complex scenarios where resolving ambiguity requires extended conversations. Those include open-ended instructions ("I'd like to hear some music") but also browsing requests like "look for some blue size 9 shoes", which could potentially involve multiple rounds of questions and answers. Second, as suggested in Section \ref{experiments}, this task can benefit from a large, diverse dataset for fine-tuning and testing models, similar to the recent dataset in \citet{lu2024weblinx} for web agents. While AndroidInteraction serves as a valuable initial benchmark for evaluating models on their performance in agent-initiated interaction in UI automation, it is not sufficiently large to facilitate comprehensive model training.

\section*{Ethical Considerations}
\paragraph{Ethical Considerations in Autonomous Agent Development}
The development of autonomous agents presents significant safety risks. One aspect of that is aligning the agent's actions with user concerns and ensuring user control. Effective agent-user interaction is crucial for mitigating these risks and enhancing the safety and usability of autonomous agents.

\paragraph{Data Collection and Privacy}
During data collection, rigorous measures were taken to protect user privacy. Annotators were explicitly instructed to mark any Personally Identifiable Information (PII), which was subsequently removed from the dataset. The dataset utilized in this research does not contain any interactions from real users.

All annotators who participated in this research provided informed consent and received fair compensation for their contributions. 

\section*{Acknowledgements}
The authors express their appreciation to Kartik Ramachandruni for their foundational work on detecting uncertainty in UI automation. We also extend our gratitude to Tapomoy Mukherjee for their leadership in guiding the annotators team collecting our data.

\clearpage

\bibliographystyle{ACM-Reference-Format}
\balance
\bibliography{bibliography}


\begin{thebibliography}{28}


\ifx \showCODEN    \undefined \def \showCODEN     #1{\unskip}     \fi
\ifx \showDOI      \undefined \def \showDOI       #1{#1}\fi
\ifx \showISBNx    \undefined \def \showISBNx     #1{\unskip}     \fi
\ifx \showISBNxiii \undefined \def \showISBNxiii  #1{\unskip}     \fi
\ifx \showISSN     \undefined \def \showISSN      #1{\unskip}     \fi
\ifx \showLCCN     \undefined \def \showLCCN      #1{\unskip}     \fi
\ifx \shownote     \undefined \def \shownote      #1{#1}          \fi
\ifx \showarticletitle \undefined \def \showarticletitle #1{#1}   \fi
\ifx \showURL      \undefined \def \showURL       {\relax}        \fi
\providecommand\bibfield[2]{#2}
\providecommand\bibinfo[2]{#2}
\providecommand\natexlab[1]{#1}
\providecommand\showeprint[2][]{arXiv:#2}

\bibitem[Bishop et~al\mbox{.}(2024)]%
        {bishop2024latent}
\bibfield{author}{\bibinfo{person}{William~E Bishop}, \bibinfo{person}{Alice Li}, \bibinfo{person}{Christopher Rawles}, {and} \bibinfo{person}{Oriana Riva}.} \bibinfo{year}{2024}\natexlab{}.
\newblock \bibinfo{title}{Latent State Estimation Helps UI Agents to Reason}.
\newblock
\newblock
\showeprint[arxiv]{2405.11120}~[cs.AI]


\bibitem[Burns et~al\mbox{.}(2022)]%
        {burns2022dataset}
\bibfield{author}{\bibinfo{person}{Andrea Burns}, \bibinfo{person}{Deniz Arsan}, \bibinfo{person}{Sanjna Agrawal}, \bibinfo{person}{Ranjitha Kumar}, \bibinfo{person}{Kate Saenko}, {and} \bibinfo{person}{Bryan~A Plummer}.} \bibinfo{year}{2022}\natexlab{}.
\newblock \showarticletitle{A dataset for interactive vision-language navigation with unknown command feasibility}. In \bibinfo{booktitle}{\emph{European Conference on Computer Vision}}. Springer, \bibinfo{pages}{312--328}.
\newblock


\bibitem[Chen et~al\mbox{.}(2017)]%
        {chen2017survey}
\bibfield{author}{\bibinfo{person}{Hongshen Chen}, \bibinfo{person}{Xiaorui Liu}, \bibinfo{person}{Dawei Yin}, {and} \bibinfo{person}{Jiliang Tang}.} \bibinfo{year}{2017}\natexlab{}.
\newblock \showarticletitle{A survey on dialogue systems: Recent advances and new frontiers}.
\newblock \bibinfo{journal}{\emph{Acm Sigkdd Explorations Newsletter}} \bibinfo{volume}{19}, \bibinfo{number}{2} (\bibinfo{year}{2017}), \bibinfo{pages}{25--35}.
\newblock


\bibitem[Deng et~al\mbox{.}(2023)]%
        {deng2023mind2web}
\bibfield{author}{\bibinfo{person}{Xiang Deng}, \bibinfo{person}{Yu Gu}, \bibinfo{person}{Boyuan Zheng}, \bibinfo{person}{Shijie Chen}, \bibinfo{person}{Samuel Stevens}, \bibinfo{person}{Boshi Wang}, \bibinfo{person}{Huan Sun}, {and} \bibinfo{person}{Yu Su}.} \bibinfo{year}{2023}\natexlab{}.
\newblock \showarticletitle{Mind2Web: Towards a Generalist Agent for the Web}.
\newblock \bibinfo{journal}{\emph{arXiv preprint arXiv:2306.06070}} (\bibinfo{year}{2023}).
\newblock


\bibitem[Hong et~al\mbox{.}(2023)]%
        {hong2023cogagent}
\bibfield{author}{\bibinfo{person}{Wenyi Hong}, \bibinfo{person}{Weihan Wang}, \bibinfo{person}{Qingsong Lv}, \bibinfo{person}{Jiazheng Xu}, \bibinfo{person}{Wenmeng Yu}, \bibinfo{person}{Junhui Ji}, \bibinfo{person}{Yan Wang}, \bibinfo{person}{Zihan Wang}, \bibinfo{person}{Yuxiao Dong}, \bibinfo{person}{Ming Ding}, {et~al\mbox{.}}} \bibinfo{year}{2023}\natexlab{}.
\newblock \showarticletitle{Cogagent: A visual language model for gui agents}.
\newblock \bibinfo{journal}{\emph{arXiv preprint arXiv:2312.08914}} (\bibinfo{year}{2023}).
\newblock


\bibitem[Kim et~al\mbox{.}(2024)]%
        {kim2024aligning}
\bibfield{author}{\bibinfo{person}{Hyuhng~Joon Kim}, \bibinfo{person}{Youna Kim}, \bibinfo{person}{Cheonbok Park}, \bibinfo{person}{Junyeob Kim}, \bibinfo{person}{Choonghyun Park}, \bibinfo{person}{Kang~Min Yoo}, \bibinfo{person}{Sang-goo Lee}, {and} \bibinfo{person}{Taeuk Kim}.} \bibinfo{year}{2024}\natexlab{}.
\newblock \showarticletitle{Aligning Language Models to Explicitly Handle Ambiguity}.
\newblock \bibinfo{journal}{\emph{arXiv preprint arXiv:2404.11972}} (\bibinfo{year}{2024}).
\newblock


\bibitem[Li et~al\mbox{.}(2025)]%
        {li2025effects}
\bibfield{author}{\bibinfo{person}{Wei Li}, \bibinfo{person}{William~E Bishop}, \bibinfo{person}{Alice Li}, \bibinfo{person}{Christopher Rawles}, \bibinfo{person}{Folawiyo Campbell-Ajala}, \bibinfo{person}{Divya Tyamagundlu}, {and} \bibinfo{person}{Oriana Riva}.} \bibinfo{year}{2025}\natexlab{}.
\newblock \showarticletitle{On the effects of data scale on ui control agents}.
\newblock \bibinfo{journal}{\emph{Advances in Neural Information Processing Systems}}  \bibinfo{volume}{37} (\bibinfo{year}{2025}), \bibinfo{pages}{92130--92154}.
\newblock


\bibitem[Li et~al\mbox{.}(2023)]%
        {li2023uinav}
\bibfield{author}{\bibinfo{person}{Wei Li}, \bibinfo{person}{Fu-Lin Hsu}, \bibinfo{person}{Will Bishop}, \bibinfo{person}{Folawiyo Campbell-Ajala}, \bibinfo{person}{Oriana Riva}, {and} \bibinfo{person}{Max Lin}.} \bibinfo{year}{2023}\natexlab{}.
\newblock \showarticletitle{UINav: A maker of UI automation agents}.
\newblock \bibinfo{journal}{\emph{arXiv preprint arXiv:2312.10170}} (\bibinfo{year}{2023}).
\newblock


\bibitem[Li et~al\mbox{.}(2020)]%
        {li2020mapping}
\bibfield{author}{\bibinfo{person}{Yang Li}, \bibinfo{person}{Jiacong He}, \bibinfo{person}{Xin Zhou}, \bibinfo{person}{Yuan Zhang}, {and} \bibinfo{person}{Jason Baldridge}.} \bibinfo{year}{2020}\natexlab{}.
\newblock \showarticletitle{Mapping natural language instructions to mobile UI action sequences}.
\newblock \bibinfo{journal}{\emph{arXiv preprint arXiv:2005.03776}} (\bibinfo{year}{2020}).
\newblock


\bibitem[Li et~al\mbox{.}(2024)]%
        {li2024personal}
\bibfield{author}{\bibinfo{person}{Yuanchun Li}, \bibinfo{person}{Hao Wen}, \bibinfo{person}{Weijun Wang}, \bibinfo{person}{Xiangyu Li}, \bibinfo{person}{Yizhen Yuan}, \bibinfo{person}{Guohong Liu}, \bibinfo{person}{Jiacheng Liu}, \bibinfo{person}{Wenxing Xu}, \bibinfo{person}{Xiang Wang}, \bibinfo{person}{Yi Sun}, {et~al\mbox{.}}} \bibinfo{year}{2024}\natexlab{}.
\newblock \showarticletitle{Personal llm agents: Insights and survey about the capability, efficiency and security}.
\newblock \bibinfo{journal}{\emph{arXiv preprint arXiv:2401.05459}} (\bibinfo{year}{2024}).
\newblock


\bibitem[L{\`u} et~al\mbox{.}(2024)]%
        {lu2024weblinx}
\bibfield{author}{\bibinfo{person}{Xing~Han L{\`u}}, \bibinfo{person}{Zden{\v{e}}k Kasner}, {and} \bibinfo{person}{Siva Reddy}.} \bibinfo{year}{2024}\natexlab{}.
\newblock \showarticletitle{Weblinx: Real-world website navigation with multi-turn dialogue}.
\newblock \bibinfo{journal}{\emph{arXiv preprint arXiv:2402.05930}} (\bibinfo{year}{2024}).
\newblock


\bibitem[Rawles et~al\mbox{.}(2023)]%
        {rawles2023android}
\bibfield{author}{\bibinfo{person}{Christopher Rawles}, \bibinfo{person}{Alice Li}, \bibinfo{person}{Daniel Rodriguez}, \bibinfo{person}{Oriana Riva}, {and} \bibinfo{person}{Timothy Lillicrap}.} \bibinfo{year}{2023}\natexlab{}.
\newblock \showarticletitle{Android in the wild: A large-scale dataset for android device control}.
\newblock \bibinfo{journal}{\emph{arXiv preprint arXiv:2307.10088}} (\bibinfo{year}{2023}).
\newblock


\bibitem[Reid et~al\mbox{.}(2024)]%
        {reid2024gemini}
\bibfield{author}{\bibinfo{person}{Machel Reid}, \bibinfo{person}{Nikolay Savinov}, \bibinfo{person}{Denis Teplyashin}, \bibinfo{person}{Dmitry Lepikhin}, \bibinfo{person}{Timothy Lillicrap}, \bibinfo{person}{Jean-baptiste Alayrac}, \bibinfo{person}{Radu Soricut}, \bibinfo{person}{Angeliki Lazaridou}, \bibinfo{person}{Orhan Firat}, \bibinfo{person}{Julian Schrittwieser}, {et~al\mbox{.}}} \bibinfo{year}{2024}\natexlab{}.
\newblock \showarticletitle{Gemini 1.5: Unlocking multimodal understanding across millions of tokens of context}.
\newblock \bibinfo{journal}{\emph{arXiv preprint arXiv:2403.05530}} (\bibinfo{year}{2024}).
\newblock


\bibitem[Ren et~al\mbox{.}(2023)]%
        {ren2023robots}
\bibfield{author}{\bibinfo{person}{Allen~Z Ren}, \bibinfo{person}{Anushri Dixit}, \bibinfo{person}{Alexandra Bodrova}, \bibinfo{person}{Sumeet Singh}, \bibinfo{person}{Stephen Tu}, \bibinfo{person}{Noah Brown}, \bibinfo{person}{Peng Xu}, \bibinfo{person}{Leila Takayama}, \bibinfo{person}{Fei Xia}, \bibinfo{person}{Jake Varley}, {et~al\mbox{.}}} \bibinfo{year}{2023}\natexlab{}.
\newblock \showarticletitle{Robots that ask for help: Uncertainty alignment for large language model planners}.
\newblock \bibinfo{journal}{\emph{arXiv preprint arXiv:2307.01928}} (\bibinfo{year}{2023}).
\newblock


\bibitem[Schneider et~al\mbox{.}(2022)]%
        {schneider2022mobile}
\bibfield{author}{\bibinfo{person}{Stefan Schneider}, \bibinfo{person}{Stefan Werner}, \bibinfo{person}{Ramin Khalili}, \bibinfo{person}{Artur Hecker}, {and} \bibinfo{person}{Holger Karl}.} \bibinfo{year}{2022}\natexlab{}.
\newblock \showarticletitle{mobile-env: An open platform for reinforcement learning in wireless mobile networks}. In \bibinfo{booktitle}{\emph{NOMS 2022-2022 IEEE/IFIP Network Operations and Management Symposium}}. IEEE, \bibinfo{pages}{1--3}.
\newblock


\bibitem[Sun et~al\mbox{.}(2022)]%
        {sun2022meta}
\bibfield{author}{\bibinfo{person}{Liangtai Sun}, \bibinfo{person}{Xingyu Chen}, \bibinfo{person}{Lu Chen}, \bibinfo{person}{Tianle Dai}, \bibinfo{person}{Zichen Zhu}, {and} \bibinfo{person}{Kai Yu}.} \bibinfo{year}{2022}\natexlab{}.
\newblock \showarticletitle{META-GUI: towards multi-modal conversational agents on mobile GUI}.
\newblock \bibinfo{journal}{\emph{arXiv preprint arXiv:2205.11029}} (\bibinfo{year}{2022}).
\newblock


\bibitem[Todi et~al\mbox{.}(2021)]%
        {todi2021conversations}
\bibfield{author}{\bibinfo{person}{Kashyap Todi}, \bibinfo{person}{Luis~A Leiva}, \bibinfo{person}{Daniel Buschek}, \bibinfo{person}{Pin Tian}, {and} \bibinfo{person}{Antti Oulasvirta}.} \bibinfo{year}{2021}\natexlab{}.
\newblock \showarticletitle{Conversations with GUIs}. In \bibinfo{booktitle}{\emph{Proceedings of the 2021 ACM Designing Interactive Systems Conference}}. \bibinfo{pages}{1447--1457}.
\newblock


\bibitem[Toyama et~al\mbox{.}(2021)]%
        {toyama2021androidenv}
\bibfield{author}{\bibinfo{person}{Daniel Toyama}, \bibinfo{person}{Philippe Hamel}, \bibinfo{person}{Anita Gergely}, \bibinfo{person}{Gheorghe Comanici}, \bibinfo{person}{Amelia Glaese}, \bibinfo{person}{Zafarali Ahmed}, \bibinfo{person}{Tyler Jackson}, \bibinfo{person}{Shibl Mourad}, {and} \bibinfo{person}{Doina Precup}.} \bibinfo{year}{2021}\natexlab{}.
\newblock \showarticletitle{Androidenv: A reinforcement learning platform for android}.
\newblock \bibinfo{journal}{\emph{arXiv preprint arXiv:2105.13231}} (\bibinfo{year}{2021}).
\newblock


\bibitem[Venkatesh et~al\mbox{.}(2022)]%
        {venkatesh2022ugif}
\bibfield{author}{\bibinfo{person}{Sagar~Gubbi Venkatesh}, \bibinfo{person}{Partha Talukdar}, {and} \bibinfo{person}{Srini Narayanan}.} \bibinfo{year}{2022}\natexlab{}.
\newblock \showarticletitle{UGIF: UI Grounded Instruction Following}.
\newblock \bibinfo{journal}{\emph{arXiv preprint arXiv:2211.07615}} (\bibinfo{year}{2022}).
\newblock


\bibitem[Wang et~al\mbox{.}(2023)]%
        {wang2023enabling}
\bibfield{author}{\bibinfo{person}{Bryan Wang}, \bibinfo{person}{Gang Li}, {and} \bibinfo{person}{Yang Li}.} \bibinfo{year}{2023}\natexlab{}.
\newblock \showarticletitle{Enabling conversational interaction with mobile ui using large language models}. In \bibinfo{booktitle}{\emph{Proceedings of the 2023 CHI Conference on Human Factors in Computing Systems}}. \bibinfo{pages}{1--17}.
\newblock


\bibitem[Wang et~al\mbox{.}(2021)]%
        {wang2021screen2words}
\bibfield{author}{\bibinfo{person}{Bryan Wang}, \bibinfo{person}{Gang Li}, \bibinfo{person}{Xin Zhou}, \bibinfo{person}{Zhourong Chen}, \bibinfo{person}{Tovi Grossman}, {and} \bibinfo{person}{Yang Li}.} \bibinfo{year}{2021}\natexlab{}.
\newblock \showarticletitle{Screen2words: Automatic mobile UI summarization with multimodal learning}. In \bibinfo{booktitle}{\emph{The 34th Annual ACM Symposium on User Interface Software and Technology}}. \bibinfo{pages}{498--510}.
\newblock


\bibitem[Wei et~al\mbox{.}(2022)]%
        {wei2022chain}
\bibfield{author}{\bibinfo{person}{Jason Wei}, \bibinfo{person}{Xuezhi Wang}, \bibinfo{person}{Dale Schuurmans}, \bibinfo{person}{Maarten Bosma}, \bibinfo{person}{Fei Xia}, \bibinfo{person}{Ed Chi}, \bibinfo{person}{Quoc~V Le}, \bibinfo{person}{Denny Zhou}, {et~al\mbox{.}}} \bibinfo{year}{2022}\natexlab{}.
\newblock \showarticletitle{Chain-of-thought prompting elicits reasoning in large language models}.
\newblock \bibinfo{journal}{\emph{Advances in neural information processing systems}}  \bibinfo{volume}{35} (\bibinfo{year}{2022}), \bibinfo{pages}{24824--24837}.
\newblock


\bibitem[Wen et~al\mbox{.}(2023)]%
        {wen2023empowering}
\bibfield{author}{\bibinfo{person}{Hao Wen}, \bibinfo{person}{Yuanchun Li}, \bibinfo{person}{Guohong Liu}, \bibinfo{person}{Shanhui Zhao}, \bibinfo{person}{Tao Yu}, \bibinfo{person}{Toby Jia-Jun Li}, \bibinfo{person}{Shiqi Jiang}, \bibinfo{person}{Yunhao Liu}, \bibinfo{person}{Yaqin Zhang}, {and} \bibinfo{person}{Yunxin Liu}.} \bibinfo{year}{2023}\natexlab{}.
\newblock \showarticletitle{Empowering llm to use smartphone for intelligent task automation}.
\newblock \bibinfo{journal}{\emph{arXiv preprint arXiv:2308.15272}} (\bibinfo{year}{2023}).
\newblock


\bibitem[Yan et~al\mbox{.}(2023)]%
        {yan2023gpt}
\bibfield{author}{\bibinfo{person}{An Yan}, \bibinfo{person}{Zhengyuan Yang}, \bibinfo{person}{Wanrong Zhu}, \bibinfo{person}{Kevin Lin}, \bibinfo{person}{Linjie Li}, \bibinfo{person}{Jianfeng Wang}, \bibinfo{person}{Jianwei Yang}, \bibinfo{person}{Yiwu Zhong}, \bibinfo{person}{Julian McAuley}, \bibinfo{person}{Jianfeng Gao}, {et~al\mbox{.}}} \bibinfo{year}{2023}\natexlab{}.
\newblock \showarticletitle{Gpt-4v in wonderland: Large multimodal models for zero-shot smartphone gui navigation}.
\newblock \bibinfo{journal}{\emph{arXiv preprint arXiv:2311.07562}} (\bibinfo{year}{2023}).
\newblock


\bibitem[Yang et~al\mbox{.}(2023)]%
        {yang2023appagent}
\bibfield{author}{\bibinfo{person}{Zhao Yang}, \bibinfo{person}{Jiaxuan Liu}, \bibinfo{person}{Yucheng Han}, \bibinfo{person}{Xin Chen}, \bibinfo{person}{Zebiao Huang}, \bibinfo{person}{Bin Fu}, {and} \bibinfo{person}{Gang Yu}.} \bibinfo{year}{2023}\natexlab{}.
\newblock \showarticletitle{Appagent: Multimodal agents as smartphone users}.
\newblock \bibinfo{journal}{\emph{arXiv preprint arXiv:2312.13771}} (\bibinfo{year}{2023}).
\newblock


\bibitem[Zhan and Zhang(2023)]%
        {zhan2023you}
\bibfield{author}{\bibinfo{person}{Zhuosheng Zhan} {and} \bibinfo{person}{Aston Zhang}.} \bibinfo{year}{2023}\natexlab{}.
\newblock \showarticletitle{You only look at screens: Multimodal chain-of-action agents}.
\newblock \bibinfo{journal}{\emph{arXiv preprint arXiv:2309.11436}} (\bibinfo{year}{2023}).
\newblock


\bibitem[Zhang et~al\mbox{.}(2021)]%
        {zhang2021screen}
\bibfield{author}{\bibinfo{person}{Xiaoyi Zhang}, \bibinfo{person}{Lilian De~Greef}, \bibinfo{person}{Amanda Swearngin}, \bibinfo{person}{Samuel White}, \bibinfo{person}{Kyle Murray}, \bibinfo{person}{Lisa Yu}, \bibinfo{person}{Qi Shan}, \bibinfo{person}{Jeffrey Nichols}, \bibinfo{person}{Jason Wu}, \bibinfo{person}{Chris Fleizach}, {et~al\mbox{.}}} \bibinfo{year}{2021}\natexlab{}.
\newblock \showarticletitle{Screen recognition: Creating accessibility metadata for mobile applications from pixels}. In \bibinfo{booktitle}{\emph{Proceedings of the 2021 CHI Conference on Human Factors in Computing Systems}}. \bibinfo{pages}{1--15}.
\newblock


\bibitem[Zhang et~al\mbox{.}(2020)]%
        {zhang2020recent}
\bibfield{author}{\bibinfo{person}{Zheng Zhang}, \bibinfo{person}{Ryuichi Takanobu}, \bibinfo{person}{Qi Zhu}, \bibinfo{person}{MinLie Huang}, {and} \bibinfo{person}{XiaoYan Zhu}.} \bibinfo{year}{2020}\natexlab{}.
\newblock \showarticletitle{Recent advances and challenges in task-oriented dialog systems}.
\newblock \bibinfo{journal}{\emph{Science China Technological Sciences}} \bibinfo{volume}{63}, \bibinfo{number}{10} (\bibinfo{year}{2020}), \bibinfo{pages}{2011--2027}.
\newblock


\end{thebibliography}

\appendix

\section{Model Error Analysis} \label{app:error_analysis}
An analysis of the discrepancies between the gold labels and the predictions of our top-performing models (text and text+screen, both two-stage with few-shot prompting) reveals several patterns. Most discrepancies are attributed to model errors, though a small proportion (15\% for text+screen, 26\% for text-only) can be considered acceptable or subjective, particularly in the case of false negatives. This suggests that the models tend to miss interactions that could be omitted without impacting task completion, rather than generating unnecessary interactions that annotators missed.

Screen understanding emerges as the primary source of false negatives in both modalities, with the incorporation of screenshots not significantly improving this aspect. For instance, the models often struggle when scrolling is needed to reveal relevant UI options. Notably, the text-only model frequently misses cases where personal details are required, a weakness overcome by the inclusion of visual input.

Regarding false positives, the most prevalent issue for both models is generating unnecessary confirmations (52\% of false positives for the text-only model, 51\% for the text+screen model). This is followed by navigation-related questions in the multimodal model (31\%), and suggestions to perform follow-up tasks in the text-only model (21\%). These findings highlight areas for improvement in model design and training, particularly in enhancing screen understanding and reducing unnecessary interactions.

\section{Additional Examples from AndroidInteraction} \label{app:additional_examples}
We present here several examples from the AndroidControl dataset that our annotation process identified as requiring user interaction. For each example we provide the original screenshots and action descriptions for all the steps up to the point where interaction was deemed necessary. The corresponding user instruction and the our suggested message to the user were given in the captions. 

\begin{figure*}[t]
    \centering
    \includegraphics[width=0.6\textwidth,height=5.5cm]{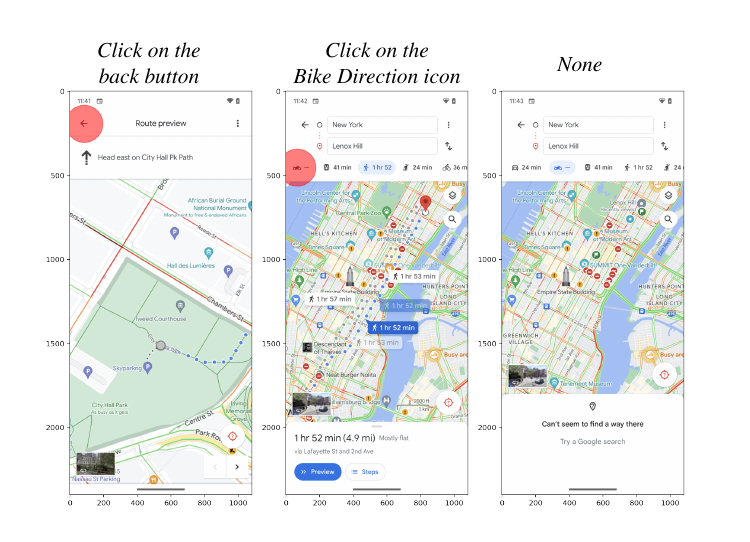}
    \caption{Example episode from the AndroidControl dataset, with user instruction "I want biking directions as I am getting late to reach my destination". The last step was annotated as requiring user interaction in AndroidInteraction, with the message: "Seems like Bike route is not available for this journey".}
    \Description{Screenshots from an episode showing a user switching to "biking" travel mode for a specific route on Google Maps. The last screen shows the app was unable to find a suitable route.}
\end{figure*}

\begin{figure*}[t]
    \centering
    \includegraphics[width=0.7\textwidth,height=5.5cm]{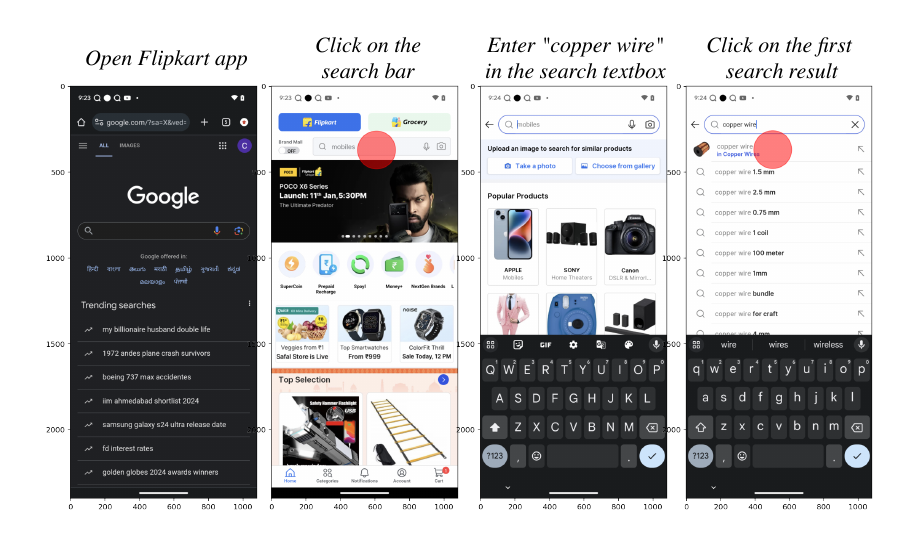}
    \caption{Example episode from the AndroidControl dataset, with user instruction "As I have a short circuit, I'd want to search for copper wire Round in the Flipkart app.". The last step was annotated as requiring user interaction in AndroidInteraction, with the message: "What type of Copper Wire you are looking for?".}
    \Description{Screenshots from an episode showing a user searching for a copper wire while e-shopping. The last screen shows there is a category of copper wires in the app, alongside different specific gauges of wires matching the user query.}
\end{figure*}

\begin{figure*}[t]
    \centering
    \includegraphics[width=0.55\textwidth,height=5.5cm]{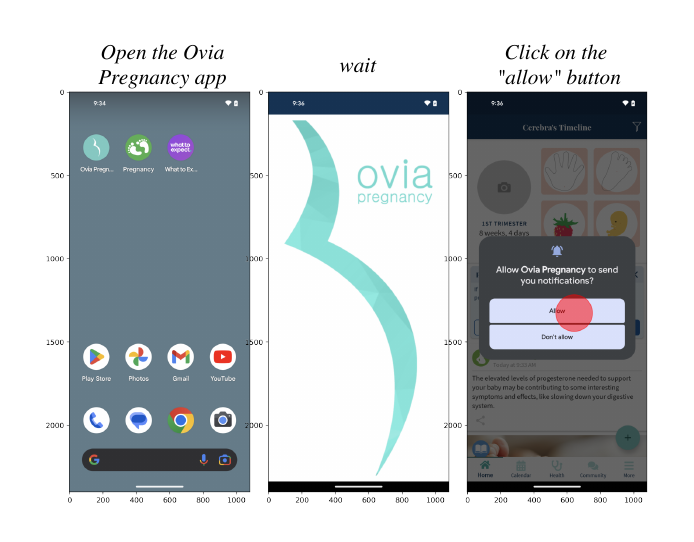}
    \caption{Example episode from the AndroidControl dataset, with user instruction "Open the Ovia Pregnancy app \& turn on the Bump tracker filter.". The last step was annotated as requiring user interaction in AndroidInteraction, with the message: "Would you like to allow Ovia Pregnancy to send you notifications?".}
    \Description{Screenshots from an episode showing a user opening a pregnancy app, and a pop-up asking to allow app notifications in the final screenshot.}
\end{figure*}

\end{document}